\documentclass[aps,pra,showpacs,twocolumn,groupedaddress]{revtex4}

\usepackage{graphicx,bbm,amsthm, amsmath, amssymb}
\usepackage{color}

\def\tr{\,{\rm tr}\,}
\def\ket#1{|#1\rangle}
\def\bra#1{\langle#1|}
\def\braket#1#2{\langle #1 | #2 \rangle}

\def\ie{{i.e.}}

\def\ii{{\rm i}}
\def\vec#1{\underline{#1}}

\newcommand{\mm}[1]{{\mathbf{#1}}}
\def\etal#1{ {\em et al.}}
\def\tit#1{}

\newcommand{\xket}[1]{{\vert #1 \rangle}}
\newcommand{\xbra}[1]{{\langle #1 \vert}}
\newcommand{\xbraket}[2]{\langle #1 \vert #2 \rangle}
\newcommand{\cc}{ {\hat c} }
\newcommand{\ee}{ e }
\newcommand{\Aa}{ {\hat{a}} }
\newcommand{\Bb}{ {\hat{b}} }
\newcommand{\Hh}{ {\hat{\mathcal{H}}} }
\def\one{\underline{1}}

\def\one{\mathbbm{1}}

\begin{document}

\title{Complexity of thermal states in quantum spin chains}

\author{Marko \v Znidari\v c, Toma\v{z} Prosen and Iztok Pi\v{z}orn}
\affiliation{Department of physics, FMF, University of Ljubljana,
Jadranska 19, SI-1000 Ljubljana, Slovenia}

\date{\today}

\begin{abstract}
We study quantum correlations and complexity of simulation, characterized  by
quantum mutual information and entanglement entropy in operator space respectively, 
for thermal states in critical, non-critical and quantum chaotic spin chains. 
A simple general relation between the two quantities is proposed.
We show that in {\em all cases} mutual information and entanglement entropy saturate 
with the system size, whereas as a function of the {\em inverse temperature}, we find 
{\em logarithmic divergences} for critical cases
and uniform bounds in non-critical cases. A simple efficient quasi-exact method
for computation of arbitrary entropy related quantities in 
thermalized XY spin chains is proposed.
\end{abstract}


\maketitle

\section{Introduction}

Physical properties of non-integrable strongly correlated quantum systems are notoriously difficult 
to calculate due to exponential growth of Hilbert space dimension with the system size.
Nevertheless, many numerical and analytical approaches have been developed that provide results for 
special properties and specific systems. The question arises, which quantities are amenable to classical calculation in spite of generic system complexity?

Recently, two results emerging from the field of quantum information shed light on the question. First, it has been shown that for one-dimensional (1D) systems the entanglement content of ground states grows logarithmically with the number of particles for fermionic critical systems 
and saturates for non-critical or bosonic systems \cite{werner,latorre,Cardy,amico}.
Second, {\em density matrix renormalization group} (DMRG) method \cite{white} has been
reinterpreted and optimized within {\em matrix product state} (MPS) ansatz 
\cite{MPS,MPO} for the many body wave-function. 
Its complexity is essentially given by the {\em entanglement entropy} of 
the system's bipartition.
Despite being essentially the best method available it is not 
known precisely in which instances is DMRG computationally efficient, for example it has been shown to
be {\em inefficient} for computing the evolution of {\em homogeneous} 1D non-integrable systems in 
{\em real} time \cite{PRE} but becomes efficient in the presence of {\em disorder} due to the emergence of (many body) localization \cite{ZPP}. Thermal states at finite temperature have been 
simulated using an extension of MPS to density operators ({\em matrix product operators} MPO), and by performing the
evolution in {\em imaginary time} (inverse temperature) \cite{MPO}. 
However, no analysis of efficiency scaling with the system size $n$ and inverse temperature 
$\beta$ have been performed. Note that finding a ground state -- corresponding to the limit 
$\beta\to\infty$ -- is proven to be a {\em hard} problem \cite{jens},
even though it can often be performed efficiently \cite{white}.
For a different algorithm for computing thermal states see Ref.~\cite{Hastings:07}. 
In Ref.~\cite{Hastings:06} an upper bound on the necessary MPO dimension is found for thermal states which 
unfortunately scales exponentially with $\beta$. It is not clear if one can do better, that is,  
if finite temperature states can be calculated efficiently, uniformly
in $\beta$? In the present work we show that this is indeed possible.

Complexity will be characterized by the {\em operator space entanglement entropy} (OSEE) 
\cite{prosenPRA76} of thermal states which is directly linked to the minimal necessary
rank of MPO ansatz. OSEE is defined as von Neumann entropy of the reduced density matrix
defined with respect to a physical density operator treated as an element of
a Hilbert (Fock) space of operators.
 In addition we shall show that OSEE behaves in the same way as the
{\em quantum mutual information} (QMI) $I_{A:B}$~\cite{Groisman:05}, which is a natural measure of quantum
correlations in a bipartition $A+B$ and an {\em upper bound} for distillable entanglement.
Our main result is that OSEE and $I_{A:B}$ in 
thermodynamic limit $n\to\infty$ grow as $\sim (c/3)\log_2\beta$ for critical (gapless) systems
and {\em saturate} for non-critical (gapped) systems,
while they {\em always saturate} with $\beta$ for a finite $n$.
This shows that the linear in $\beta$ upper bound on QMI of Ref.\cite{Wolf:08} is 
far from optimal and DMRG computations of large (infinite) 
critical 1D quantum systems should be {\em efficient} at {\em all nonzero} temperatures.
This has to be contrasted with typical {\em linear} growth of OSEE in $t$ in {\em real time}
evolution \cite{PRE}.
Note that OSEE and QMI always saturate with $n$, for any fixed $\beta$, and hence
scale very differently from thermodynamic entropy of a block of length $n$,
which for a critical system behaves as \cite{Korepin:04} 
 $S \propto \frac{n}{\beta}$ for large $n$.
Our results are based on analytical and numerical calculations in several examples of
integrable and non-integrable spin chains. We outline a novel method for quasi-exact calculations of
OSEE and $I_{A:B}$ in XY spin $1/2$ chains using the concept of Fock space of operators 
\cite{prosen3Q}, whereas we rely on numerical MPO-DMRG simulations for the other cases. 

\section{Methods}
First, we describe how to construct a thermal state and compute 
OSEE in a general spin chain, regardless of its integrability, using MPO 
ansatz and imaginary time evolution. A thermal state 
$\rho(\beta)=\exp{(-\beta H)}/\tr{\left[\exp{(-\beta H)}\right]}$
of Hamiltonian $H$ of a chain of $n$ spins $1/2$
can be expressed as an element of $4^n$ dimensional Hilbert space ${\cal K}$,
$\ket{\rho} = \sum_{\vec{s}}  c_{\vec{s}} \ket{\sigma^{\vec{s}}}$
spanned by Pauli operators $\sigma^{\vec{s}}\equiv \sigma_1^{s_1} \cdots \sigma_{n}^{s_{n}}$, 
$\vec{s}\equiv s_1,\ldots,s_n$, $s_m\in\{0,1,2,3\}$,
$\sigma^0\equiv\mathbbm{1},\sigma^{1,2,3}\equiv\sigma^{x,y,z}$. Inner product in ${\cal K}$ is
defined as $\braket{x}{y}=2^{-n}\tr x^\dagger y$.
Within MPO ansatz expansion coefficients $c_{\vec{s}}$ 
are represented in terms of $4n$ $D\times D$ matrices $\mm{A}_m^s$, $m=1,\ldots,n$, $s=0,1,2,3$, as
\begin{equation}
c_{\vec{s}} = \tr{(\mm{A}_1^{s_1} \cdots \mm{A}_{n}^{s_{n}})}.
\label{eq:MPO}
\end{equation}
For exact MPO representation matrix dimension of $\mm{A}^s_m$ must be equal to the number of 
nonzero Schmidt coefficients for a bipartite cut of a vector $c_{\vec{s}}$ at the corresponding site. 
Although for generic density operators $\rho$ the required dimension $D$ is exponentially large 
(in $n$) in practice one truncates the matrices to a smaller dimension $D$ and uses 
(\ref{eq:MPO}) as a variational ansatz.
As a rough estimate of a minimal $D \sim 2^{S^\sharp}$, which is a direct measure of {\em complexity} 
of $\rho$, one can use OSEE $S^\sharp$ defined as an {\em entanglement entropy} 
of a super-ket 
$\ket{\rho}$
\begin{equation}
S^\sharp=-\tr_{\! A}{(\mm{R}\,\log_2{\mm{R}})},\qquad \mm{R}=
\braket{\rho}{\rho}^{-1} \tr_{\! B}{ \ket{\rho} \bra{\rho}}
\label{eq:S}
\end{equation}
where a subscript $A$ denotes the first $n_A$ sites and $B$ its complement.
More detailed analysis \cite{schuch} specifies exact criteria for the applicability of matrix product ansatz (\ref{eq:MPO}) in terms of properties of the moments of Schmidt coefficients.
Writing the expansion coefficients $c_{\vec{s}}$ as a $2^{n_A}\times 2^{n-n_A}$ dimensional matrix 
$C_{\vec{s}_A,\vec{s}_B}=c_{\vec{s}_A \vec{s}_B}$, the reduced density matrix
(\ref{eq:S}) is given by $\mm{R}=\mm{C} \mm{C}^T/\tr{(\mm{C} \mm{C}^T)}$ and
quantifies the bipartite correlations and the difficulty of a 
MPO simulation of a given density operator $\rho$. For use of OSEE in a non-MPO context see Ref.~\cite{operator}.
$S^\sharp$ is directly related to the entanglement entropy of physical states only in the case of pure 
states, $\rho=\ket{\psi}\bra{\psi}$, in which case $S^\sharp$ is just twice the 
entanglement entropy of a pure state $\psi$, e.g. the ground state at zero temperature. 
However, even though $S^\sharp$ is not an entanglement monotone for general $\rho$ it can be used
as a bipartite entanglement estimator for thermal states $\rho(\beta)$. Namely, as we will demonstrate
later, its scaling on $n$ and $\beta$ is the same 
as that of QMI $I_{A:B}$, defined as
\begin{equation}
I_{A:B}=S(\rho_A)+S(\rho_B)-S(\rho),
\label{eq:I}
\end{equation}
where $\rho_A=\tr_{\! B}\rho$, $\rho_B=\tr_{\! A}\rho$, and 
$S(\sigma)=-\!\tr{(\sigma\log_2{\sigma})}$. 
QMI measures total correlations between $A$ and $B$ and is an upper bound
for relative entropy of entanglement which in turn is an upper bound for 
distillable entanglement $E_D(\rho)$~\cite{Vedral:98}. MPO representation for a thermal state 
$\rho(\beta)$, apart from normalization, 
is computed by performing imaginary time evolution in $\beta$, 
$\rho(\beta+\epsilon) = \exp(-\epsilon H/2) \rho(\beta) \exp(-\epsilon H/2)$,
starting at an infinite temperature state 
$\rho(0) = \mathbbm{1}/2^n$
having $D=1$.
Such $\beta$-evolution increases OSEE and as a consequence, dimension 
$D$ has to increase as well in order to preserve the accuracy.
Assuming the empirical fact that a {\em fixed} small $\epsilon$ (in our case $\epsilon=0.05$) is
enough to produce accurate results in a {\em finite} number of steps
(for all $\beta <\infty$), we conclude that time complexity of simulation is
polynomial in $D$ and $\beta$. 
Note that in contrast to a (unitary) {\em real time} evolution, imaginary time evolution 
does not preserve canonical Schmidt decomposition structure of matrices 
$\mm{A}_m^{s_m}$ \cite{MPS}. 
In order to improve stability we apply local rotations every few steps in 
order to ``reorthogonalize'' the matrices $\mm{A}_m^{s_m}$.
\\\\
Quasi-exact calculation of $S^\sharp$ and $I_{A:B}$ for thermal states is possible
for integrable chains which can be mapped to quadratic fermionic models by Wigner-Jordan transformation, e.g. 1D XY model. 
Most general such situation is described by a Hamiltonian
$H = \underline{w} \cdot \mathbf{H} \underline{w} \equiv \sum_{j,l=1}^{2n} H_{jl} w_j w_l$
where $\mm{H}$ is an {\em anti-symmetric} {\em Hermitian} 
$2n\times 2n$ matrix and
$w_{2m-1} = \sigma_m^x \prod_{l=1}^{m-1} \sigma_l^z$,
$w_{2m} = \sigma_m^y \prod_{l=1}^{m-1} \sigma_l^z$ are  Majorana operators
satisfying $\{w_j,w_l\}\equiv w_j w_l + w_l w_j =2\delta_{jl}$.
It is crucial to recognize \cite{prosen3Q, prosenPRA76} that ${\cal K}$ is in fact a 
{\em Fock space} if spanned by $4^n$ products of $w_j$, 
$P_{\vec{\alpha}} = w_1^{\alpha_1}\cdots w_{2n}^{\alpha_{2n}}$,
$\alpha_j\in\{0,1\}$, and equipped with adjoint Fermi maps 
$\cc_j \ket{P_{\vec{\alpha}}} = \delta_{\alpha_j,1}\ket{w_j P_{\vec{\alpha}}}$.
In our notation $\hat{\bullet}$ designates a linear map over the Fock space of operators
${\cal K}$. Note that OSEE is just an entanglement entropy of an element of ${\cal K}$. 
Therefore, using the formalism of \cite{latorre} lifted to an operator space,
one can calculate $S^\sharp$ from the {\em anti-symmetric} correlation matrix $\mm{\Gamma}$ defined by
\begin{equation}
\frac{\xbra{\ee^{-\beta H}} \Aa_p \Aa_q \xket{\ee^{-\beta H}}}{ \xbraket{\ee^{-\beta H}}{\ee^{-\beta H}}} = \delta_{pq} + \ii \Gamma_{pq}, \quad p,q=1,\ldots,4n
\label{eq:Gamma1}
\end{equation}
where $\Aa_{2j-1}\equiv (\cc_j+\cc^\dagger_j)$,
$\Aa_{2j}\equiv \ii (\cc_j-\cc^\dagger_j)$, $j=1,\ldots,2n$ satisfying $\{\Aa_p,\Aa_q\}=2\delta_{pq}$.
Thermal state in (\ref{eq:Gamma1}) can be written in terms of dynamics
\begin{equation}
\ket{e^{-\beta H}} = \exp(-\beta \Hh) \ket{\one}
\label{eq:thdyn}
\end{equation}
using a quadratic adjoint Hamiltonian map $\Hh$ over ${\cal K}$, 
$\Hh = (\vec{\cc}+\vec{\cc}^\dagger)\cdot \mathbf{H}
       (\vec{\cc}+\vec{\cc}^\dagger)$,
since $(\cc_j+\cc^\dagger_j)\ket{\rho} = \ket{w_j \rho}$.
Diagonalizing the matrix $\mathbf{H}$ we obtain eigenvalues 
$\lambda_k,-\lambda_k$, ordered such that $0\le \lambda_1\le \ldots \le \lambda_n$, and the corresponding {\em orthonormal} eigenvectors $\underline{v}_k, \underline{v}_k^*$ in terms of which 
we write the adjoint Hamiltonian in (\ref{eq:thdyn}) in the 
{\em normal form} $\Hh = \sum_{k=1}^{n} 4\lambda_{k}(\Bb_{k}^\dagger \Bb_{k}-\tfrac{1}{2})$ where
$\Bb_k=\underline{v}_{k}^*\cdot(\vec{\cc}+\vec{\cc}^\dagger)/\sqrt{2}$.
Identities
$\xbra{\one} \Bb_{k} \xket{\one}=0$, 
 $\xbra{\one}\Bb_{k}\Bb_{l}\xket{\one}=0$, 
$\xbra{\one} \Bb_{k}^\dagger \Bb_{l} \xket{\one} = \frac{1}{2}\delta_{kl}$
allow us to employ the Wick theorem to evaluate (\ref{eq:Gamma1}) 
\begin{align}
 &\Gamma_{2j-1,2l-1} = -\Gamma_{2j,2l}=-2 \sum_{k=1}^{n} \tanh(4\beta\lambda_k)\textrm{Im}(v_{k,j}^*v_{k,l}^{}) 
 \nonumber \\
 &\Gamma_{2j,2l-1} = \delta_{jl} -2 \sum_{k=1}^{n} 
\bigl(1-\frac{1}{\cosh(4\beta\lambda_k)}\bigr)\textrm{Re}(v_{k,j}^*v_{k,l}^{}).
\label{eq:Gammaresult}
\end{align}
Finally, OSEE is obtained from the eigenvalues 
$\pm\ii \nu_j$ of $4n_A\times 4n_A$ upper-left block of $\mathbf{\Gamma}$ as
$S^\sharp = \sum_{j=1}^{2n_A} H_2(\frac{1+\nu_j}{2})$ where 
$H_2(x)=-x\log_2 x-(1-x)\log_2(1-x)$ 
\cite{latorre}.  

Similarly, we can also easily compute the block entropy $S(\rho_A)$ of a 
thermal state $\rho(\beta)$  by diagonalizing 
$H = \sum_{k=1}^{n} 4\lambda_k(b_k^\dagger b_k-\frac{1}{2})$ with operators 
$b_k = \underline{v}_k^* \cdot \underline{w}/\sqrt{2}$ (which are elements of
${\cal K}$ and not maps over ${\cal K}$ like $\Bb_k$) 
and computing the correlation matrix $\mm{\Gamma}'$, defined as
$\tr (w_j w_l e^{-\beta H})/\tr e^{-\beta H} = \delta_{jl} + \ii \Gamma'_{jl}$, $j,l=1,\ldots,2n$. 
Expressing $w_j$ and $H$ in terms of $b_k,b_k^\dagger$, and using the Wick theorem with
$\tr (b_k^\dagger b_l) = \frac{1}{2}\delta_{kl} \tr\one $ results in a general 2-point thermal 
correlation function
\begin{equation}
 \Gamma'_{jl} = -2\sum_{k=1}^{n} \tanh(2\beta\lambda_{k})\textrm{Im}(v_{k,j}^*v_{k,l}^{}). 
\label{eq:Gamma2}
\end{equation}
Diagonalizing $2n_A\times 2n_A$ block of $\mathbf{\Gamma}'$ 
corresponding to the sub-lattice $A$, 
yielding eigenvalues $\pm \ii \nu'_j$, we obtain the block entropy 
$S(\rho_A) = \sum_{j=1}^{n_A} H_2(\frac{1+\nu_j'}{2})$.
Considering also $B$, and $A+B$ instead of $A$ we get QMI (\ref{eq:I}).

\section{Results}
We shall show results for OSEE and QMI for several families of 
critical and non-critical as well as integrable and non-integrable quantum spin chains
\begin{align}
H_{XY}&=\sum_{l=1}^{n-1} \left(\frac{1\!+\!\gamma}{2}\sigma_l^x \sigma_{l+1}^x+\frac{1\!-\!\gamma}{2}\sigma_l^y \sigma_{l+1}^y \right) + \sum_{l=1}^{n} h\sigma_l^z,
\nonumber\\
H_{I}&=\sum_{l=1}^{n-1}\sigma_l^x \sigma_{l+1}^x+\sum_{l=1}^{n} (h_x \sigma_l^x+h_z \sigma_l^z),
\label{eq:Hamil}\\
H_{XXZ}&=\sum_{l=1}^{n-1}(\sigma_l^x\sigma_{l+1}^x+\sigma_l^y\sigma_{l+1}^y+\Delta\sigma_l^z\sigma_{l+1}^z)+\sum_{l=1}^{n} h_l\sigma_l^z,
\nonumber
\end{align} 
all for open boundary conditions,
and a symmetric cut at $n_A=n/2$ where largest bipartite entanglement is expected.
Majorana representation of the integrable 
XY case which is necessary for explicit
calculations [Eqs. (\ref{eq:Gamma1}-\ref{eq:Gamma2})] is
$H_{XY}=-\ii\sum_{l=1}^{n-1}(\frac{1+\gamma}{2}w_{2l}w_{2l+1}-\frac{1-\gamma}{2}w_{2l-1}w_{2l+2})
-\ii\sum_{l=1}^n h w_{2l-1}w_{2l}.
$

\begin{figure}[h!]
\centerline{\includegraphics[width=3.3in]{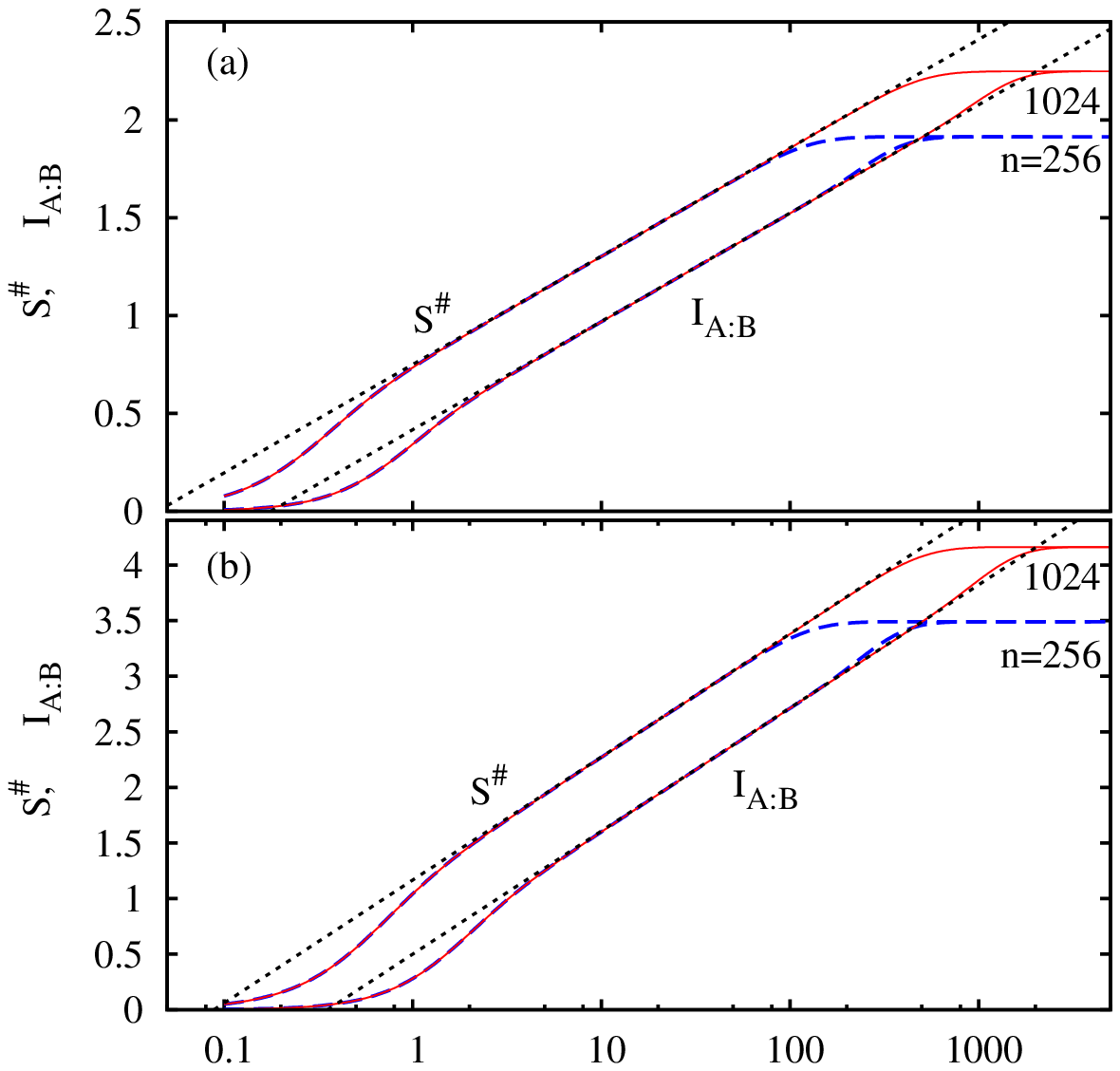}}
\vspace{-7mm}
\centerline{\includegraphics[width=3.3in]{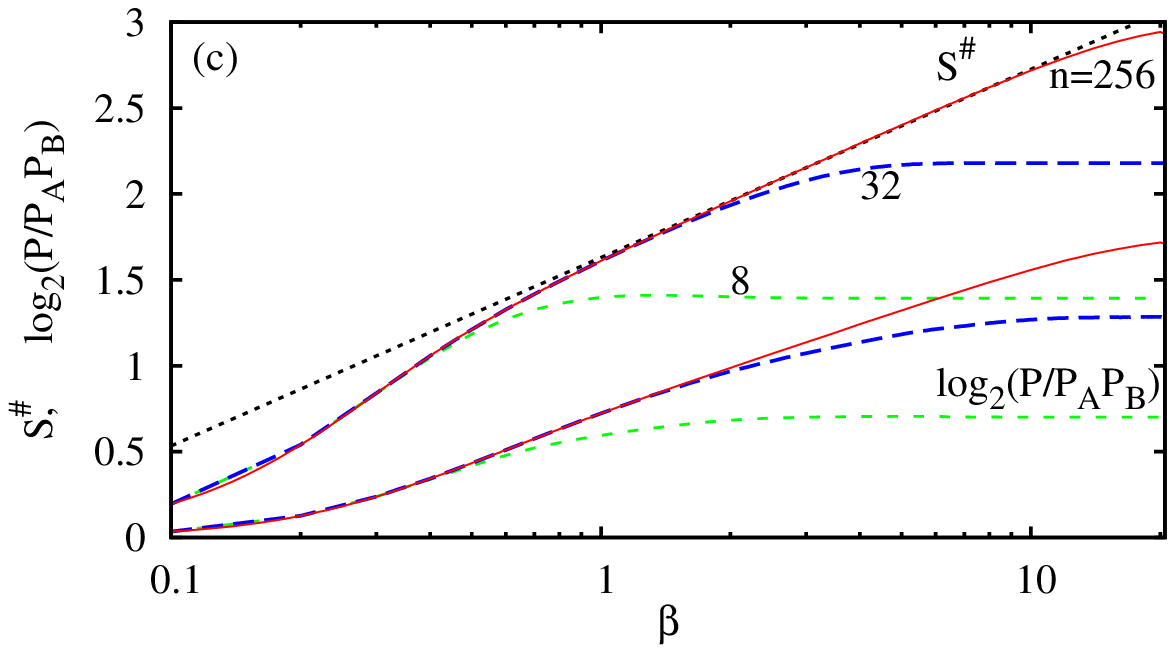}}
\caption{Logarithmic growth of OSEE $S^\sharp$ (upper curves) and QMI $I_{A:B}$ (lower curves) for 
thermal states of 
critical models: (a) XX chain $H_{XY}(\gamma=0,h=0)$, 
(b) transverse Ising chain $H_{XY}(\gamma=1,h=1)$, both for $n=1024,256$,
and (c) XXZ chain $H_{XXZ}(\Delta=0.5,h_l\equiv 0)$ for $n=256,32,8$.
In case (c) we show mutual purity $I_P$ instead of $I_{A:B}$. 
Dotted lines indicate conjectures (\ref{eq:conj}), and (\ref{eq:conj2}) combined with
 (\ref{eq:conj}) [see text].
}
\label{fig:crit}
\end{figure}

In Fig.~\ref{fig:crit} we show the results for {\em quantum critical} systems for which the entanglement 
entropy of the ground state grows as $S_0 \sim (c/3)\log_2 n$~\cite{latorre} where $c$ is the central charge
of the corresponding conformal field theory \cite{Cardy}. However, we 
take several very large $n$ and 
plot $\beta$ dependence of OSEE and QMI, for (a) XX model $H_{XY}(\gamma=0,h=0)$, (b)
transverse Ising model $H_{XY}(\gamma=1,h=1)$, and (c)
XXZ model $H_{XXZ}(\Delta=0.5,h_l\equiv 0)$. These critical models are all completely integrable,
however we were able to make exact calculations only for (a,b), whereas we used MPO-DMRG simulations for (c) where - since evaluating QMI (\ref{eq:I}) is computationally 
difficult even within MPO ansatz - we have instead computed {\em mutual purity},
defined as $I_P=\log_2{\left[ P(\rho)/( P(\rho_A)P(\rho_B))\right]}$ with $P(\sigma)=\tr{\sigma^2}$.
It is well known that purity often well describes the behavior of von Neumann entropy and its
complement is sometimes termed linear entropy. In addition we checked for thermal states 
of small systems that mutual purity and  QMI behave similarly. 

From Fig.~\ref{fig:crit} we conclude that in the large 
$n$ limit OSEE grows logarithmically with $\beta$
\begin{equation}
S^\sharp(\beta) = \frac{c}{3}\log_2{\beta}+c',
\label{eq:conj}
\end{equation}
with parameters $c=1$ and $c'=1.17$ for XX model (a), $c=1/2$ and $c'=0.75$ for 
transverse Ising model (b), and $c=1$ and $c'=1.63$ for XXZ model (c). $c$ is exactly the
central charge of the corresponding model.
Presently we do not have an explanation for this finding and we state it as a 
conjecture.
For fixed $n$, $S^\sharp$ saturates for $\beta$ larger than the inverse spectral gap 
$\beta^*=1/\Delta E$ at twice the ground state value $2S_0 \sim (2c/3)\log_2{n}$.

\begin{figure}[h!]
\centerline{\includegraphics[width=3.3in]{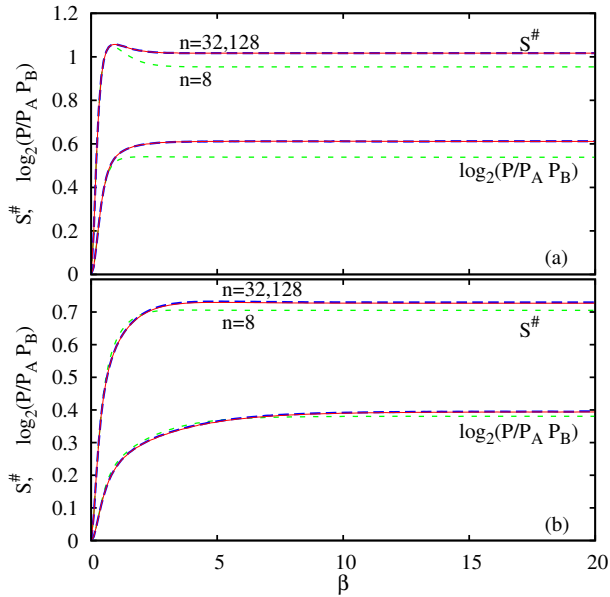}}
\caption{Saturation of OSEE $S^\sharp$ (upper curves) and mutual purity $I_P$ 
(lower curves) for quantum chaotic 
{\em noncritical} systems, namely (a) for XXZ model in a staggered field
$H_{XXZ}(\Delta=0.5,h_l=-(1+(-1)^l)/2)$ and (b) for tilted Ising 
$H_I(h_x=1,h_z=1)$, for $n=8,32,128$. Note that the curves for $n=32$ and $n=128$
are practically indistinguishable.}
\label{fig:gapped}
\end{figure}

\begin{figure}[t!]
\centerline{\includegraphics[width=3.3in]{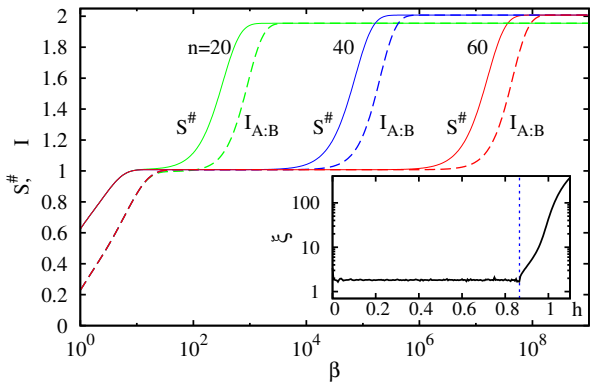}}
\caption{OSEE $S^{\sharp}$ (full curves) and QMI $I_{A:B}$ (broken curves) for
{\em noncritical} XY chain with $\gamma=0.5$, $h=0.9$, and $n=20,40,60$. Note a crossover between two
saturation plateaus due to a single exponentially close (in $n$) excited state. 
The inset shows $h$ dependence of the gap decay length
$\xi$, computed from the fit to 
 $\Delta E \propto e^{-n/\xi}$, for $n=30,\ldots,60$. Vertical line indicates
 $h^*=\sqrt{1-\gamma^2}$.
 }
\label{fig:noncriticalxy}
\end{figure}

Next, let us have a look at {\em non-critical} systems. First, we will chose two {\em non-integrable}
spin models that display typical signatures of quantum chaos (Fig.~\ref{fig:gapped}): (a) Heisenberg XXZ model in a {\em staggered} magnetic field 
$H_{XXZ}(\Delta=0.5,h_l=-(1+(-1)^l)/2)$, and 
(b) Ising model in a tilted magnetic 
field $H_I(h_x=1,h_z=1)$.
We can see that for non-critical systems which possess a finite energy gap $\Delta E$ between the ground and first excited states, 
OSEE {\em saturates} at $n$ and $\beta$ independent values for sufficiently large 
$\beta \gg \beta^*$,
so the classical simulation of quantum thermal states for gapped systems is even more efficient 
than for critical ones. Although this might seem expected, one should know that
for OSEE we do not have theoretical uniform bounds in $n$ like for QMI 
\cite{Wolf:08}.
Note that in the presence of weakly broken symmetries,
the gap $\Delta E$ between two lowest states can get exponentially small with $n$ even though the system is not critical.  This for instance happens in $XY$ model.
In Fig.~\ref{fig:noncriticalxy} we show results for a non-critical integrable 
XY model (\ref{eq:Hamil}) with $\gamma=0.5$ and $h=0.9$, where $\beta^*=1/\Delta E$ 
grows exponentially with $n$. 
Actually we have shown for any $H_{XY}$,  
that $\Delta E = 4\lambda_1$ where $\lambda_1$ is the smallest eigenvalue
of matrix $\mm{H}$ (\ref{eq:thdyn}), and which in turn has been found to decay exponentially 
$\lambda_1 \sim \exp(-n/\xi)$. Interestingly, for 
$h^2 + \gamma^2 < 1$,
the gap decay length $\xi$ is practically {\em insensitive} to parameters 
$\gamma,h$, whereas at the point
$h^* = \sqrt{1-\gamma^2}$, where the ground state is doubly degenerate
and separable \cite{kurmann}, $\xi$ instantly starts increasing with increasing $h-h^*$ (inset of Fig.~\ref{fig:noncriticalxy}). 
However, there is no phase transition at $h^*$.

In all cases studied QMI behaves in essentially the same way as OSEE. 
In fact, for models solvable by Wigner-Jordan transformation 
$I_{A:B}(\beta)$ almost overlaps with $S^\sharp$
evaluated at $\beta/4$
\begin{equation}
I_{A:B}(\beta) \approx S^\sharp(\beta/4) 
\label{eq:conj2}
\end{equation}
apart from finite size effects at small and large $\beta$ (Figs.~\ref{fig:crit},\ref{fig:noncriticalxy}). 
We conjecture that (\ref{eq:conj2}) becomes asymptotically exact for large $n$.
Factor $4$ may be understood from comparing the formalisms of Majorana 
operators $w_j$ and the adjoint maps $\cc_j$.
Since QMI is an upper bound for distillable entanglement~\cite{Vedral:98} this means that 
exact quantum bipartite entanglement for thermal states is also small, \ie , it is either upper bounded 
by $n$ and $\beta$-independent value for non-critical systems (Figs.~\ref{fig:gapped},\ref{fig:noncriticalxy}) 
or it is upper bounded by $n$-independent value $\sim \log_2{\beta}$ for critical systems.  

\section{Conclusions}
By studying several typical models, we have demonstrated that classical simulations of thermal 
states of interacting quantum systems in 1D are {\em efficient}. Using  DMRG simulations 
combined with exact calculations for XY models we found that
in non-critical systems complexity and bipartite entanglement measures are 
uniformly bounded, both in system size $n$ and inverse temperature $\beta$,
whereas they exhibit a universal logarithmic divergence in inverse temperature
for quantum critical systems.

Our results represent a {\em uniform} finite temperature extension 
of the ground state {\em area-law}~\cite{werner,latorre,Wolf:08}.
Specifically, for critical systems, logarithmic divergence of ground state
 block entropy  with the size of the block is reflected in logarithmic divergence in 
 inverse temperature $\beta$ of the bipartite entanglement of thermal (mixed) states in the thermodynamic limit. Even though our results are not rigorous they suggest optimal bounds on entanglement and complexity of thermal states that are qualitatively stronger than 
previously known \cite{Wolf:08}.
 
We thank J. Eisert for useful comments and acknowledge support by the grants P1-0044 and J1-7347 of 
Slovenian Research Agency.

\end{document}